# Coupled-Mode Theory for Stationary and Nonstationary Resonant Sound Propagation

Theodoros T. Koutserimpas and Romain Fleury*

*Institute of Electrical Engineering, École Polytechnique Fédérale de Lausanne (EPFL), 1015 Lausanne, Switzerland*

*To whom correspondence should be addressed. Email: romain.fleury@epfl.ch

## Abstract

We present a complete analytical derivation of the equations used for stationary and nonstationary wave systems regarding resonant sound transmission and reflection described by the phenomenological Coupled-Mode Theory. We calculate the propagating and coupling parameters used in Coupled-Mode Theory directly by utilizing the generalized eigenwave-eigenvalue problem from the Hamiltonian of the sound wave equations. This Hamiltonian formalization can be very useful since it has the ability to describe mathematically a broad range of acoustic wave phenomena. We demonstrate how to use this theory as a basis for perturbative analysis of more complex resonant scattering scenarios. In particular, we also form the effective Hamiltonian and coupled-mode parameters for the study of sound resonators with background moving media. Finally, we provide a comparison between Coupled-Mode theory and full-wave numerical examples, which validate the Hamiltonian approach as a relevant model to compute the scattering characteristics of waves by complex resonant systems.

# I. Introduction

Wave manipulation and control by resonant media has been on the frontline of wave engineering research in the past decades. Research in wave physics and material science has illustrated extreme wave propagation and scattering properties, which enrich the theory of wave physics and involuntary oscillations, and can be applied to a broad spectrum of potential applications and technologies. From metamaterials and metasurfaces to photonic and phononic crystal devices, resonators have attested to play a crucial role to impart extraordinary wave propagation properties and effective parameters. Coupled-Mode Theory (CMT) is a mathematical abstract tool employed to formulate the energy transitions of coupled resonant systems. CMT is mainly used to predict traveling waves or oscillating states in classical and quantum mechanics, quantum electronics, electromagnetics and acoustics [1–11]. More specifically, it has been used to describe wave propagation properties in stationary, nonstationary, linear and nonlinear resonances [1,12–16]. Due to its simplicity, the resulting equations of CMT attract a lot of attention and usage by researchers. While the simplicity of such method is undeniable, the parameters used are either a product of intuition of the examined physical system or a result of experimental or numerical fitting procedures. In this research paper, we consider the case of resonant sound wave transmission and reflection and we provide the exact analytical coupled-mode-theory parameters *without* presuming specific wave conditions or implementing numerical fitting techniques. This is achieved by formulating a Schrödinger type matrix formalization of the acoustic wave equations regarding the sound pressure and the particle velocity. The power of this approach is that it can serve as a basis to study the scattering of perturbed resonant systems from the coupled-mode theory of the unperturbed resonator. We evidence this by considering the example of resonant sound transmission and reflection in nonstationary systems, i.e. acoustic resonators subject to the motion of the acoustic background medium. The CMT analysis and the results are in total agreement with the full-wave simulations, which applied the finite element method (FEM) [17].

## II. The Generalized Eigenvalue Problem: From Sound Waves to Schrödinger Formalization

The equations regarding linear acoustical wave propagation in a lossless homogenous medium can take the simple form of

$$\nabla \vec{u} = -\frac{1}{K_0}\partial_t p, \tag{1}$$

$$\nabla p = -\rho_0 \partial_t \vec{u}, \tag{2}$$

where $\rho_0$ is the density, $K_0$ is the bulk modulus of the medium and $\vec{u}, p$ are the particle velocity and the sound pressure, respectively. Note that $c = \sqrt{K_0/\rho_0}$ where $c$ is the velocity of the acoustic wave in the medium. Equations (1) and (2) can form a Schrödinger system with the appropriate mathematical manipulation of the equations. For this purpose, we define the sound state function

$$|\Psi\rangle = \begin{pmatrix} p \\ \vec{u} \end{pmatrix}. \tag{3}$$

The inner product of such vectors is defined as $\langle \Psi_a | \Psi_b \rangle = \iiint \left( p_a^* p_b + \vec{u}_a^* \vec{u}_b \right) dV$. We also define a weighting operator related to the acoustical properties of the medium

$$\hat{\zeta} = \begin{pmatrix} \frac{1}{K_0} & 0 \\ 0 & \rho_0 \mathbf{I}_{3\times 3} \end{pmatrix}. \tag{4}$$

In general $\hat{\zeta}$ can describe any type of medium (isotropic, anisotropic and bianisotropic), however here we focus on the simple case of a homogenous fluid. Notice that the assumption of lossless medium results in a Hermitian weighting operator $(\hat{\zeta} = \hat{\zeta}^{\dagger})$. The Hamiltonian of the system is given by

$$\mathcal{H} = -j \begin{pmatrix} 0 & \nabla \cdot \\ \nabla & 0 \end{pmatrix}. \tag{5}$$

Notice that, by definition, the Hamiltonian is Hermitian if and only if $\langle \Psi_a | \mathcal{H} \Psi_b \rangle = \langle \mathcal{H} \Psi_a | \Psi_b \rangle$ for two arbitrary solutions of the same boundary-condition problem. It is straightforward to prove that the Hamiltonian is indeed Hermitian when the system is made of finite resonances (the fields attenuate and the integrals become zero for $r \to \infty$) or when the system is periodic and the sound state functions have to obey the Bloch theorem [18]. Finally, by manipulating Equations (3)-(5), it is overt that the resulting Schrödinger system is

$$j \hat{\zeta} \cdot \partial_t | \Psi \rangle = \mathcal{H} | \Psi \rangle. \tag{6}$$

If one considers the case of time-harmonic fields the problem is simplified. The sound state function becomes $|\Psi(\vec{r},t)\rangle = |\psi(\vec{r},t)\rangle e^{-j\omega t}$, and the Schrödinger equation (6) takes the form

$$\omega_m \hat{\zeta} | \psi_m \rangle = \mathcal{H} | \psi_m \rangle. \tag{7}$$

For Hermitian matrices $\mathcal{H}$ and $\hat{\zeta}$ the resonant frequencies $\omega_m$ are real. Different eigenfrequencies $\omega_m \neq \omega_n$ correspond to eigenwaves $|\psi_m\rangle, |\psi_n\rangle$ which are orthogonal. Hence, they satisfy the following orthogonality condition

$$\langle \psi_n | \hat{\zeta} \psi_m \rangle = 0. \tag{8}$$

In addition, it is easily shown that in such Hermitian systems the existence of a solution $|\psi_m\rangle$ with positive (or negative) frequency leads to the formulation of an extra independent solution $|\hat{\sigma}_z \psi_m\rangle$ with negative (or positive) frequency, i.e.:

$$-\omega_m \hat{\zeta} | \hat{\sigma}_z \psi_m \rangle = \mathcal{H} | \hat{\sigma}_z \psi_m \rangle, \tag{9}$$

where $\hat{\sigma}_z$ is the third order Pauli matrix. In order to return the solution of (9) to the same frequency with (7) we can simply apply the complex conjugation operator. By doing so, we obtain that the system has two independent solutions at the same frequency. Such mathematical result can be interpreted as the fact that in any homogenous reciprocal medium a duality of wave solutions with opposite propagating direction is always found.

$$|\psi_m^+\rangle = \begin{pmatrix} p_m \\ \vec{u}_m \end{pmatrix}, \quad |\psi_m^-\rangle = \begin{pmatrix} p_m^* \\ -\vec{u}_m^* \end{pmatrix}. \tag{10}$$

Taking this into account, the orthogonality relation (8) has to hold:

$$\langle \psi_n | \hat{\zeta} \hat{\sigma}_z \psi_m \rangle = 0. \tag{11}$$

The Equation (11) is of great physical importance, since in addition to the obvious orthogonality that it describes, it also proves that for the same mode ($\omega_m = \omega_n$) the potential energy of the wave mode which is stored in the acoustic pressure is the same with the kinetic energy stored in the motion of the medium's particles $\left( \iiint \frac{1}{K_0} |p_m|^2 \, dV = \iiint \rho_0 |\vec{u}_m|^2 \, dV \right)$.

### III. Eigenwave Bases and Scattering Analysis

#### A. Traveling Wave Basis

We now assume a traveling wave basis for the solution of the wave distribution at some frequency near a resonant frequency $\omega_m$. This wave formulation assumes

$$|\Psi\rangle = \alpha_+(\vec{r},t) |\psi_m^+\rangle e^{-j\omega_m t} + \alpha_-(\vec{r},t) |\psi_m^-\rangle e^{-j\omega_m t}. \tag{12}$$

To move further, we need to make important approximations in (12) about the variations of the envelopes of the waves in positive and negative direction, $\alpha_+$ and $\alpha_-$. These envelopes are presumed to be slowly varying in space and time. Note also that modes $|\psi_m^\pm\rangle$ are assumed to be invariant under space derivatives, since all the information about the wave's spatial distribution variations are defined in the envelopes $\alpha_\pm$. Plugging (12) to (6) gives the following equation

$$j\hat{\zeta}\left(\partial_t \alpha_+ |\psi_m^+\rangle + \partial_t \alpha_- |\psi_m^-\rangle\right) = \mathcal{H}(\alpha_+)|\psi_m^+\rangle + \mathcal{H}(\alpha_-)|\psi_m^-\rangle, \qquad (13)$$

where $\mathcal{H}(\alpha_\pm)$ is the Hamiltonian operating just on the envelopes $\alpha_\pm$. Eq. (13) provides the fundamental relation for forming the CMT. It is possible to form CMT by applying the ket $\langle\psi_m^\pm|$ to Eq. (13). Taking into account that:

$$\langle\psi_m^-|\hat{\zeta}\psi_m^+\rangle = 2\iiint \frac{1}{K_0} p_m^2 dV \qquad (14)$$

$$\langle\psi_m^+|\hat{\zeta}\psi_m^-\rangle = 2\iiint \frac{1}{K_0} \left(p_m^*\right)^2 dV \qquad (15)$$

$$\langle\psi_m^\pm|\hat{\zeta}\psi_m^\pm\rangle = 2\iiint \frac{1}{K_0} |p_m|^2 dV \qquad (16)$$

and the system's overall directional energy flux is

$$\vec{\mathcal{F}}_e = \iiint \mathrm{Re}\{p_m^* \vec{u}_m\} dV, \qquad (17)$$

the resulting two projected equations are

$$\partial_t \alpha_+ + (\vec{v}_g \nabla)\alpha_+ = -g^* \partial_t \alpha_-, \qquad (18)$$

$$\partial_t \alpha_- - (\vec{v}_g \nabla)\alpha_- = -g \partial_t \alpha_+, \qquad (19)$$

where $g = \frac{\langle \psi_m^- | \hat{\zeta} \psi_m^+ \rangle}{\langle \psi_m^\pm | \hat{\zeta} \psi_m^\pm \rangle}$, and $\vec{v}_g = \frac{2\vec{\mathcal{F}}_e}{\langle \psi_m^\pm | \hat{\zeta} \psi_m^\pm \rangle}$. Examining closer these parameters it is easy to extract some interesting features. For example, the parameter $g$ can characterize the homogeneity of the medium and the localization of the field in the resonators. If the chosen medium has no defects or cavities $\langle \psi_m^- | \hat{\zeta} \psi_m^+ \rangle \approx 0$, and in consequence $g \approx 0$, which means that there is no coupling between the two envelopes $\alpha_+, \alpha_-$. But if the scatterer supports a mode then the field is localized and $g \neq 0$. Of course, the absolute value of $g$ has an upper bound for any case of wave propagation, due to the way it is normalized $(|g| \leq 1)$. Taking a closer look at $\vec{v}_g$, we find that aside from the normalization, it is proportional to the vector $\vec{S} = \text{Re}\{p_m \vec{u}_m^*\}$, which describes the acoustic power flow. This directional flux satisfies the solenoidal field condition: $\nabla \vec{S} = 0$ in steady state and in the absence of active elements, due to the general continuity equation. The resulting algebra for the computation of $\vec{v}_g$ provides a quantitative parameter that characterizes the effective group velocity of the wave. An additional remark is that the inverse norm, i.e. $1/|\vec{v}_g|$ is proportional to the decay rate of the resonator. All these quantities can be computed from the knowledge of the mode profiles $p_m$ and $\vec{u}_m$.

**B. Standing Wave Basis**

Instead of applying the traveling waves $|\psi_m^+\rangle, |\psi_m^-\rangle$ as basis of the solution, it is possible to define standing waves $|\psi_A\rangle, |\psi_B\rangle$ which, depending on the under examined problem, could provide a more practical mathematical formulation when used as basis of the wave solution. The even $(|\psi_A\rangle)$ and odd $(|\psi_B\rangle)$ modes are defined as

$$|\psi_A\rangle = \frac{|\psi_m^+\rangle + |\psi_m^-\rangle}{2} = \begin{pmatrix} \text{Re}\{p_m\} \\ j\,\text{Im}\{\vec{u}_m\} \end{pmatrix}, \tag{20}$$

$$|\psi_B\rangle = \frac{|\psi_m^+\rangle - |\psi_m^-\rangle}{2j} = \begin{pmatrix} \text{Im}\{p_m\} \\ -j\,\text{Re}\{\vec{u}_m\} \end{pmatrix}, \tag{21}$$

These modes represent even and odd standing waves. This overt observation is confirmed mathematically since the $\vec{S}$ vector is zero separately for both $|\psi_A\rangle, |\psi_B\rangle$. It is also straightforward to construct the traveling waves from the odd and even standing bases since: $|\psi_m^\pm\rangle = |\psi_A\rangle \pm j|\psi_B\rangle$. In the case of choosing the standing wave basis we assume wave solutions of the form

$$|\Psi\rangle = A(\vec{r},t)|\psi_A\rangle e^{-j\omega_m t} + jB(\vec{r},t)|\psi_B\rangle e^{-j\omega_m t}. \tag{22}$$

Plugging Equation (22) to (6) gives us a general equation

$$j\hat{\zeta}\left(\partial_t A|\psi_A\rangle + j\partial_t B|\psi_B\rangle\right) = \mathcal{H}(A)|\psi_A\rangle + j\mathcal{H}(B)|\psi_B\rangle. \tag{23}$$

Following the same mathematical procedure as before we form the projections of both modes (even and odd) to Equation (23). This results to two Equations, which follow

$$(1+g)\partial_t A + (\vec{v}_g \nabla) B = 0, \tag{24}$$

$$(1-g)\partial_t B + (\vec{v}_g \nabla) A = 0, \tag{25}$$

where $g, \vec{v}_g$ are defined as in the previous section. This formulation based on even and odd modes can be very useful when dealing with mirror symmetric structures and centrosymmetric acoustic crystals. In such structures, it is common that either the odd or the even mode dominate, depending on the operating frequency. It is straightforward to show that $g = \dfrac{\iiint (1/K_0)\left[p_A^2 - p_B^2\right]dV}{\iiint (1/K_0)\left[p_A^2 + p_B^2\right]dV}$. In the case that the even mode dominates $g \to 1$ and the system of equations is significantly simplified. In a close analogy, when the odd mode dominates $g \to -1$.

## C. Scattering Analysis

Let us assume a one-dimensional structure from $x = 0$ to $x = L$ as depicted in Fig. 1. The notations $u_+$ and $v_+$ are used to symbolize the incoming signals on the left and the right side and $u_-, v_-$ the outgoing signals on the left and the right side, respectively. The scattering analysis comes down to the specification of the scattering matrix, i.e. the matrix $S_0$, which connects the outgoing with the incoming wave signals

$$\begin{pmatrix} u_- \\ v_- \end{pmatrix} = S_0 \begin{pmatrix} u_+ \\ v_+ \end{pmatrix}, \tag{26}$$

$$S_0 = \begin{pmatrix} r_L & t \\ t & r_R \end{pmatrix}, \tag{27}$$

where $r_L, r_R$ are the reflection coefficients from the right and the left side and $t$ is the transmission coefficient. Utilizing the analysis based on the traveling waves and approximating the wave distributions with averages we get: $\alpha_\pm = (u_\pm + v_\mp)/2$ and using finite difference approximations, we write: $\partial_x \alpha_\pm = (v_\mp - u_\pm)/L$ and $\partial_t \alpha_\pm = -j(\omega - \omega_m)(u_\pm + v_\mp)/2$. Plugging these expressions in (18), (19) yields the scattering matrix

$$S_0 = \frac{1}{1 - \frac{j(\omega - \omega_m)}{\gamma}} \begin{pmatrix} \frac{jg(\omega - \omega_m)}{\gamma} & 1 \\ 1 & \frac{jg^*(\omega - \omega_m)}{\gamma} \end{pmatrix}, \tag{28}$$

where $\gamma = \vec{v}_g \cdot \vec{n}/L$ is the (radiative) decay rate and $\vec{n}$ is the unit vector in the direction of the wave propagation. Using the standing wave basis, we get the same result. It is straightforward to show that:

$$\frac{1+g}{2} d_t A = -\gamma A + \gamma(u_+ + v_+), \tag{29}$$

$$\frac{1-g}{2} d_t B = -\gamma B + \gamma(u_+ - v_+). \tag{30}$$

For a dominant even mode ($g \to 1$) the scattering analysis is

$$d_t A = -\gamma A + \gamma(u_+ + v_+) \tag{31}$$

$$\begin{pmatrix} u_- \\ v_- \end{pmatrix} = -\begin{pmatrix} u_+ \\ v_+ \end{pmatrix} + A \begin{pmatrix} 1 \\ 1 \end{pmatrix} \tag{32}$$

For a dominant odd mode ($g \to -1$) the scattering analysis is

$$d_t B = -\gamma B + \gamma(u_+ - v_+) \tag{33}$$

$$\begin{pmatrix} u_- \\ v_- \end{pmatrix} = \begin{pmatrix} u_+ \\ v_+ \end{pmatrix} + B \begin{pmatrix} -1 \\ 1 \end{pmatrix}. \tag{34}$$

### IV. Sound Resonance in a Background Moving Medium

**A. Perturbation Method**

In sound wave engineering, the mathematical modeling of wave propagation in nonstationary media is of great importance, in a variety of applications such as noise detection, imaging and Doppler-based devices. In these specific cases, the Galilean principle of relativity is applied and only the longitudinal material disturbance affects the propagating waves. We consider a material disturbance such that it can be modeled as a background material movement of a velocity of $\vec{w}_b$. The sound wave differential equations are:

$$(\partial_t + \vec{w}_b \cdot \nabla) p + K_0 \nabla \vec{u} = 0 \tag{35}$$

$$(\partial_t + \vec{w}_b \cdot \nabla) \vec{u} + (\vec{u} \cdot \nabla) \vec{w}_b + \nabla p / \rho_0 = 0 \tag{36}$$

In order to form the CMT of such systems we apply a perturbative method and we assume that the Hamiltonian of the acoustic system has also an extra term:

$$j\hat{\zeta} \cdot \partial_t |\Psi\rangle = \mathcal{H}_{\text{eff}} |\Psi\rangle, \tag{37}$$

$$\mathcal{H}_{\text{eff}} = \mathcal{H} + \Delta\mathcal{H}_b, \tag{38}$$

where $\mathcal{H}$ is the Hamiltonian as defined in (5) and $\Delta\mathcal{H}_b$ corresponds to the extra terms added at the wave equations (35), (36) due to the nonstationary dynamics of the medium. Following the same mathematical analysis as before, it is easily derived:

$$\Delta\mathcal{H}_b = -j \begin{pmatrix} \frac{1}{K_0}(\vec{w}_b \cdot \nabla) & 0 \\ 0 & \rho_0\left((\vec{w}_b \cdot \nabla) + \mathcal{D}\vec{w}_b\right) \end{pmatrix}, \tag{39}$$

where:

$$\mathcal{D}\vec{w}_b = \begin{pmatrix} \partial_x w_{bx} & 0 & 0 \\ 0 & \partial_y w_{by} & 0 \\ 0 & 0 & \partial_z w_{bz} \end{pmatrix}. \tag{40}$$

Assuming traveling wave basis, which is the most convenient for such problems, we form the generalized eigenvalue problem from the general equation:

$$j\hat{\zeta}\left(\partial_t \alpha_+ |\psi_m^+\rangle + \partial_t \alpha_- |\psi_m^-\rangle\right) = \mathcal{H}(\alpha_+)|\psi_m^+\rangle + \Delta\mathcal{H}_b(\alpha_+)|\psi_m^+\rangle + \mathcal{H}(\alpha_-)|\psi_m^-\rangle + \Delta\mathcal{H}_b(\alpha_-)|\psi_m^-\rangle. \tag{41}$$

Following the same analysis and projecting equation (40) with $|\psi_m^+\rangle, |\psi_m^-\rangle$ we get the modified CMT equations:

$$\partial_t \alpha_+ + (\vec{v}_{g_{\text{eff}}}^+ \nabla)\alpha_+ + m_b \alpha_+ = -g^* \partial_t \alpha_- - (\vec{\theta}_b^* \nabla)\alpha_- - \kappa_b^* \alpha_-, \tag{42}$$

$$\partial_t \alpha_- - (\vec{v}_{g_{\text{eff}}}^- \nabla)\alpha_- + m_b \alpha_- = -g \partial_t \alpha_+ - (\vec{\theta}_b \nabla)\alpha_+ - \kappa_b \alpha_+, \tag{43}$$

where the extra nonstationary coefficients are obtained from the stationary modes as:

$$\vec{v}_{g_{\text{eff}}}^{\pm} = \vec{v}_g \pm 2\iiint \vec{w}_b \left(\tfrac{|p_m|^2}{K_0}\right) dV \Big/ \langle \psi_m^{\pm} | \hat{\zeta} \psi_m^{\pm} \rangle = \vec{v}_g \pm \Delta \vec{v}_b, \tag{44}$$

$$m_b = \iiint \partial_{\vec{n}} \vec{w}_b \left(\rho_0 |\vec{u}_m|^2\right) dV \Big/ \langle \psi_m^{\pm} | \hat{\zeta} \psi_m^{\pm} \rangle, \tag{45}$$

$$\vec{\theta}_b = 2\iiint \vec{w}_b \left(\tfrac{p_m^2}{K_0}\right) dV \Big/ \langle \psi_m^{\pm} | \hat{\zeta} \psi_m^{\pm} \rangle, \tag{46}$$

$$\kappa_b = \iiint \partial_{\vec{n}} \vec{w}_b \left(\rho_0 \vec{u}_m^{\,2}\right) dV \Big/ \langle \psi_m^{\pm} | \hat{\zeta} \psi_m^{\pm} \rangle. \tag{47}$$

Evidently, the coefficients $\vec{v}_{g_{\text{eff}}}^{\pm}$ and $\vec{\theta}_b$ correspond to the scattering effects of the nonstationary operative normalized group velocity, for wave propagation at the same (+) and opposite (−) direction of the disturbance $\vec{w}_b$. In addition, $m_b$ and $\kappa_b$ are normalized coefficients which represent the dynamical influence of an attenuated or an accelerated active nonstationary background medium (at the propagating coordinate direction $\vec{n}$).

**B. Eigenfrequencies of Sound Resonances with Background Movement**

Let us assume a simple one-dimensional problem, in which the sound wave has particle velocity $\vec{u} = [u_x, 0, 0]^T$ and scalar sound pressure $p$ and there is no dependence on the other coordinates $(\partial_y = \partial_z = 0)$. Taking into account the general differential equations (35), (36) which describe sound waves in background moving wave medium we assume phasors for the scalar pressure $p$ and the particle velocity vector $\vec{u}$. It is straightforward to show with minimum algebraic manipulation that for $\vec{w}_b = [w_{bx}, w_{by}, w_{bz}]^T$ the perpendicular components $(w_{by}, w_{bz})$ of the material disturbance do not affect the wave propagation. For that reason, we can assume $\vec{w}_b = [w_{bx}, 0, 0]^T$ *without* degenerating the wave solution. This mathematical observation is physically linked with the Galilean principle of relativity [19]. The two modified equations are now:

$$j\omega p + w_{bx} d_x p + K_0 d_x u_x = 0, \tag{48}$$

$$j\omega u_x + w_{bx}d_x u_x + u_x d_x w_{bx} + (1/\rho_0)d_x p = 0. \tag{49}$$

Equations (48), (49) describe the frequency domain coupled differential equations. So far, there has been no special condition or postulation of the function of the background velocity medium. To illustrate the extraordinary dynamic wave phenomena of time-modulated resonant media let us consider the simple case of uniform motion: $w_{bx}(x) = w_{bx}$. The wave equation of pressure $p$ is:

$$d_x^2 p - \left[j2\omega w_{bx}/(c^2 - w_{bx}^2)\right]d_x p + \left[\omega^2/(c^2 - w_{bx}^2)\right]p = 0. \tag{50}$$

Equation (50) is a space-harmonic oscillator with imaginary damping coefficient. Due to space invariance and linearity, we seek for solutions of the form: $p(x) = e^{\xi x}$. Plugging this form of sound pressure into (50) gives us:

$$\xi = \frac{j\omega w_{bx}}{c^2 - w_{bx}^2} \pm \sqrt{-\left[\frac{\omega w_{bx}}{(c^2 - w_{bx}^2)}\right]^2 - \frac{\omega^2}{(c^2 - w_{bx}^2)}}. \tag{51}$$

Obviously $\xi$ corresponds to the effective wave number of the propagating sound without any precondition of the relations between the speed of sound and the background medium movement velocity. Notice that the imaginary damping coefficient results to neither an attenuated nor parametrically amplified field; an evident remark since we neglected any material losses and the disturbance has a constant uniform velocity, which does not change the system's inertness. It is well known for the stationary case that a resonant frequency for the case of a one-dimensional slab occurs when $L = n\lambda/2$, where $n \in \mathbb{Z}$, $L$ is the length of the slab and $\lambda$ is the wave length. For the same geometry if we apply a background time modulation of $w_b$ the resonant frequency is redshifted (see also the Appendix):

$$f_m' = \left(1 - \frac{w_b^2}{c^2}\right)f_m. \tag{52}$$

The relation of (52) is valid for a background movement slower than the speed of sound in the medium, as shown in the Appendix. For the case of $c \leq w_b$, the relative speed of the wave as perceived in the stationary observation frame does not allow multiple scattering at the interfaces of the slab, hence the system can no longer be characterized as resonant. To illustrate this remark better imagine a traveler running in a moving walkway of an airport (as shown in Fig. 2). The speed of the traveler (without the assistance of the walkway) is $c$, whereas the speed of the walkway floor is $w_b$. The overall speed of the traveler as observed in the stationary frame is $\vec{w}_b + \vec{c}$. In the case, that $c$ is in the same direction with $w_b$ (as depicted in Fig. 2(a)) the total speed of the traveler increases in the direction of his choice ($c + w_b$). In the case, that $c$ is in the opposite direction with $w_b$ (as depicted in Fig. 2(b)) the total speed is $|w_b - c|$ in the direction of the most dominant velocity $c$ or $w_b$. In order to have a resonant system the "traveler" (whose identity symbolizes the traveling waves inside the slab) should have the ability to run back and forth the walkway so that standing waves can be composed. When $c \leq w_b$ the "traveler" has no other option than to head into the direction of the walkway floor (or stay still for $c = w_b$), despite his/her efforts to verve to the other direction. Of course, the practical and realistic applications associated with this mathematical analysis of temporal acoustic resonators consider a smaller velocity disturbance $w_b$ with respect to the speed velocity of the medium $c$.

## V. Numerical Examples

We now prove the validity of the CMT by direct comparison to numerical examples. Let us consider the problem of reflection and transmission of a water slab amid air. Water has speed velocity $c = 1498$ m/s and density $\rho_0 = 1000$ kg/m³, whereas air has $c = 343$ m/s and $\rho_0 = 1,225$

kg/m$^3$. Assume the background speed of water is zero or perpendicular to the incident wave vector, hence the Galilean principle of relativity allows us to deal with this problem as a stationary one of a standing slab of water surrounded by air, as shown in Fig. 3(a). The slab has a length of $L = 0.2$ m. Such scattering problem can be solved analytically, by considering plane waves and boundary conditions at the boundaries of the water slab and the radiative conditions at infinity [20]. The obtained analytical solutions match perfectly the prediction obtained from Coupled-Mode Theory. As depicted in Fig. 3(b) and Fig. 3(c) the resonant frequency is $f_m = 3745$ Hz, corresponding to a resonant mode occurring under the condition: $L = \lambda/2$. In addition, for the stationary lossless case the conservation of energy stands, i.e. $|r|^2 + |t|^2 = 1$, where $r$ is the reflection coefficient and $t$ is the transmission coefficient. Furthermore, we obtain the typical Lorentzian shapeline consistent with the prediction of CMT.

Now let us consider the nonstationary problem of the same geometry, for a background longitudinal disturbance $w_b$, which we assume relatively smaller than the speed velocity of waves inside water. From the mathematical analysis of Section IV we expect a redshift of resonant frequency regardless the direction of the disturbance (i.e. independent of whether the wave is in the same or opposite direction of the incident wave). In order to solve such problem, we employ the finite element method (FEM) [17] to solve the equations in the frequency domain and we compare with the solutions obtained by CMT. Our results are depicted in Fig. 4. Fig. 4(a) shows the transmission and Fig. 4(b) shows the reflection from the nonstationary water slab. We considered the stationary state and the cases $w_b = \pm 0.02c$, $w_b = \pm 0.03c$ and $w_b = \pm 0.05c$. As we can see, the results of FEM are in complete agreement with the results acquired by CMT, *which are obtained without any fitting parameter*. The resonant frequency shifts corroborate the accuracy of the formula in (52). Of course, the expected Lorentzian shapelines are also found for these cases.

We can also verify that conservation of energy is also maintained since we do not consider losses and the constant uniform velocity disturbance doesn't alter the inertia of the overall resonant system, as predicted in Section IV B.

## VI. Conclusion

In this work, we developed a simple mathematical approach based on Hamiltonian physics in order to derive analytical formulae for the CMT parameters. CMT can be utilized and is easily implemented for the solution of general problems regarding sound resonant systems and acoustic wave propagation through reflections and transmission by resonant scattering objects. Such analysis can be utilized to mathematically model the interactions of time-periodically modulated sound resonators or even nonlinear wave properties in acoustical propagation. In order to illustrate the general applicability of Coupled-Mode Theory, we developed the perturbation analysis of the Hamiltonian to deal with nonstationary acoustic resonators. We proved that in nonstationary resonant sound systems a uniform background velocity disturbance can alter the eigenfrequencies of the resonator, resulting to a redshifted eigenfrequency response. Numerical simulations employing finite element method to the background moving acoustic medium equations validate our results.

## Acknowledgment

This work was supported by the Swiss National Science Foundation (SNSF) under Grant No. 172487.

## Appendix: Nonstationary Eigenfrequency Analysis

Taking into account Equation (51) and after basic algebraic manipulations, we get:

$$\xi = \frac{j\omega w_{bx}}{c^2 - w_{bx}^2} \pm \sqrt{-\left[\frac{\omega w_{bx}}{(c^2 - w_{bx}^2)}\right]^2 - \frac{\omega^2}{(c^2 - w_{bx}^2)}} = \frac{j\omega(w_{bx} \pm c)}{c^2\left(1 - \frac{w_{bx}^2}{c^2}\right)} = j\delta \pm j\beta \quad (53)$$

Let us remind that the two solutions of $\xi$ correspond to wave propagation at $-\hat{x}$ and $+\hat{x}$ direction and are valid for any relation between the speed of sound $c$ and the background medium movement velocity $w_{bx}$. Since we study sound cavities that resonate caused by multiple scattering at the interfaces, the constant part of the wave number $j\delta$ is eliminated due to phase compensation. In more detail for the 1-D sound resonator of length $L$ the phase deviation due to $j\delta$ of the traveling wave at the $+\hat{x}$ direction results: $\phi_\delta^+ = -\delta L$, whereas at the $-\hat{x}$ direction results: $\phi_\delta^- = \delta L$. It is clear that $\phi_\delta^{tot} = \phi_\delta^+ + \phi_\delta^- = 0$. For this reason, we can easily determine the effective speed velocity (in regards to the computation of the resonant frequency). This expected value is given by the relation:

$$\beta = \frac{\omega}{c\left(1 - \frac{w_{bx}^2}{c^2}\right)} = \frac{\omega}{c_{eff}}, \quad (54)$$

where $c_{eff} = c\left(1 - \frac{w_{bx}^2}{c^2}\right)$. The consistent relations for the resonant frequencies for the stationary and nonstationary acoustic problem of the same geometry are:

$$c = \lambda f_m, \quad (55)$$

$$c_{eff} = \lambda_{eff} f'_m, \quad (56)$$

and on top of Equations (55), (56) we know that $\lambda = \lambda_{eff} = 2L/n$, where $n \in \mathbb{Z}$ for the one-dimensional sound resonator of a length of $L$, since they correspond to the resonance response of the one-dimensional sound system (similar conditions can be found for 2-D and 3-D resonators).

Dividing (56) and (55) we get the relation used directly in Equation (52) in the main text:

$$f'_m = \frac{c_{\text{eff}}}{c} f_m = \left(1 - \frac{w_{bx}^2}{c^2}\right) f_m. \qquad (57)$$

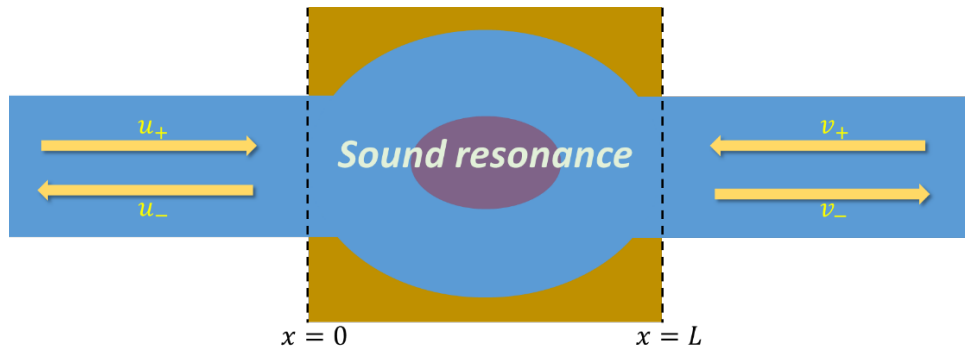

Figure 1: Schematic representation of a one-dimensional sound resonant system with input and output wave signals from the left and right directions.

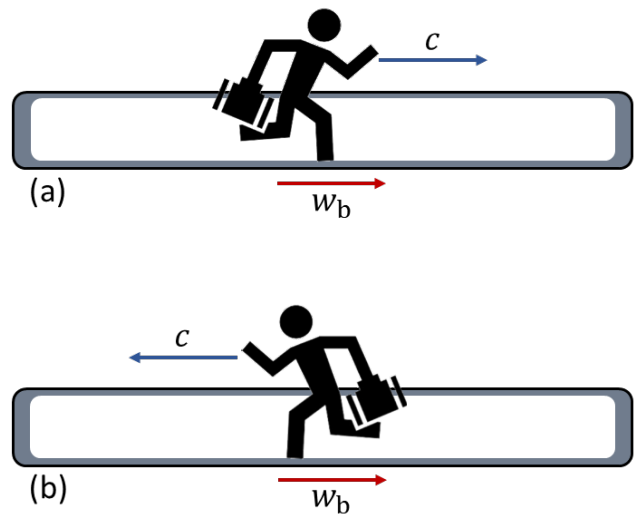

Figure 2: Traveler running in a moving walkway of an airport, (a) in the same direction as the assisting floor of the walkway, (b) in the opposite direction of the assisting floor of the walkway. If $w_b < c$, the traveler is able to run back and forth between the two sides of the walkway, whereas if $w_b \geq c$ the traveler is heading to the direction of the moving walkway even if he/she chooses to run the opposite direction, or remains still for $w_b = c$. By analogy, a resonator subject to a background medium faster than the sound speed cannot support multiple reflections and the resonance disappears.

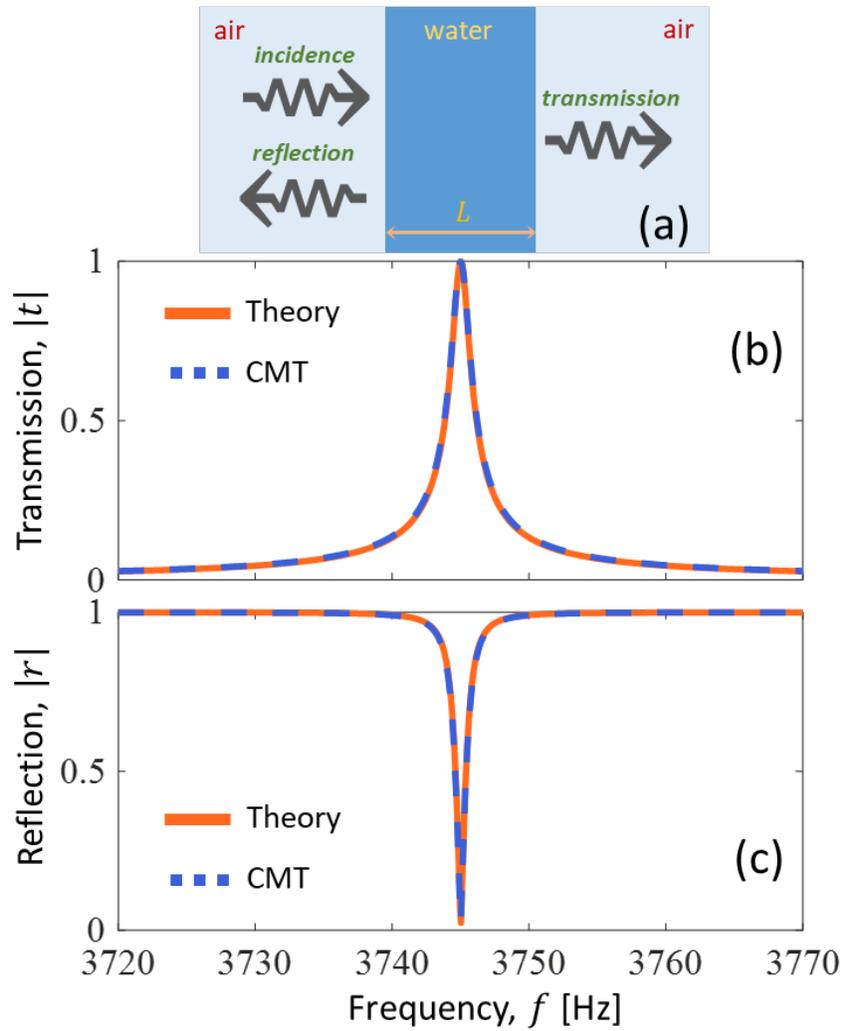

Figure 3: (a) Schematic representation of the numerical example of a 20 cm-thick water slab amid air. (b) Graphical plot and comparison of theory and CMT of the transmission coefficient and (c) graphical plot and comparison of theory and CMT of the reflection coefficient of the wave propagation problem under examination.

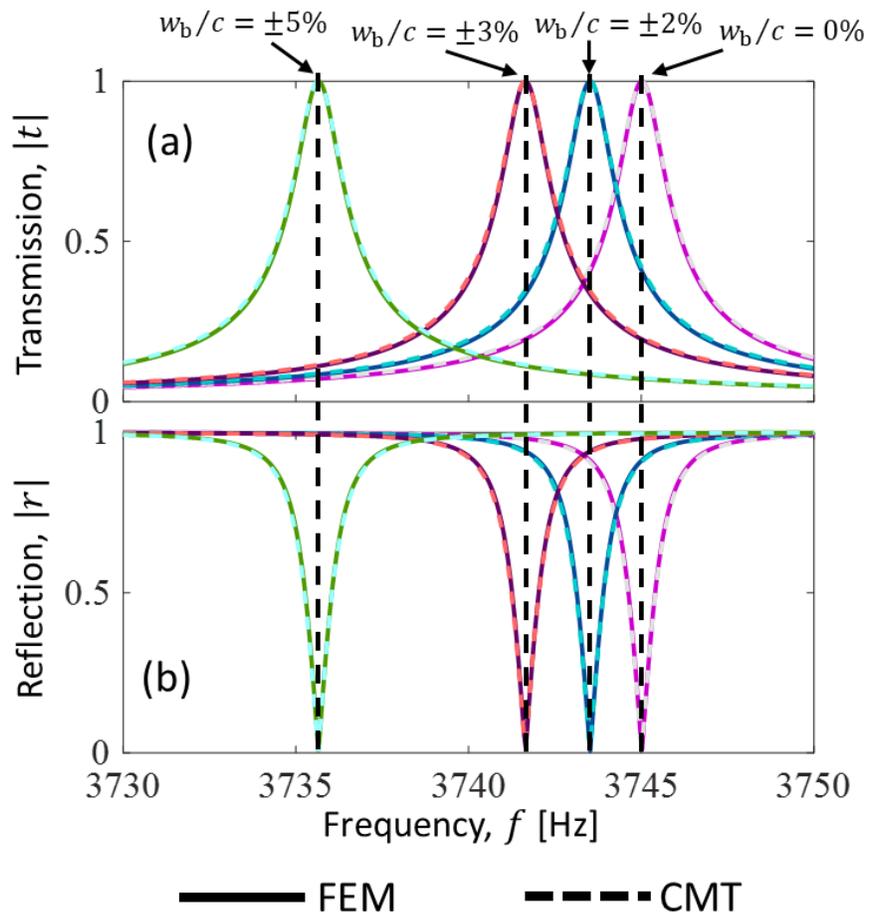

Figure 4: Graphical plots of FEM and CMT results of nonstationary sound resonators for: $w_b = 0$, $w_b = \pm 0.02c$, $w_b = \pm 0.03c$ and $w_b = \pm 0.05c$, where $c$ is the speed of sound in the slab. Panel (a) portrays the transmission coefficient and (b) the reflection coefficient. Full-wave simulations and CMT agree perfectly without using any fitting parameter.